\documentclass[amssymb,prb,twocolumn]{revtex4}
\usepackage{graphicx}
\newfont{\bl}{cmbxsl10 scaled\magstep1}
\begin{document}
\draft
\title{Interlayer Exchange Coupling Mediated by Valence Band Electrons}
\author{J. Blinowski}
\affiliation{ Institute of Theoretical Physics, Warsaw University, ul.~Ho\.za 69,00-681 Warszawa, Poland}
\author{P. Kacman}
\affiliation{Institute of Physics, Polish Academy of Sciences, al.~Lotnik\'ow 32/46, 02-668 Warszawa, Poland}
\date{\today}
\begin{abstract}

The interlayer exchange coupling mediated by valence band electrons in
all-semiconductor IV-VI magnetic/nonmagnetic superlattices is studied
theoretically. A 3D tight-binding model, accounting for the band and
magnetic structure of the constituent superlattice components is used to
calculate the spin-dependent part of the total electronic energy.  The
antiferromagnetic coupling between ferromagnetic layers in EuS/PbS
superlattices is obtained, in agreement with the experimental evidences.
The results obtained for the coupling between antiferromagnetic layers in
EuTe/PbTe superlattices are also presented.
\end{abstract}
\maketitle
  
\section{Introduction}
Interlayer exchange coupling (IEC) was discovered in late 80-ties in
Fe/Cr/Fe trilayers \cite{gruen}. Since then it has been observed in a
variety of multilayer structures composed of alternating magnetic and
nonmagnetic layers. These studies concentrated on the coupling between
ferromagnetic, metallic layers, although separated by both metallic and
insulating spacers. Thus, the ferromagnetic character of the magnetic
layers and the fact that in these structures the Fermi level was situated
in the region of high density of electronic states inhered to the
theoretical models, which were designed to explain the origins of the IEC
phenomena (see \cite{bruno} and the references therein).  Surprisingly
enough, the IEC was also discovered in all-semiconductor superlattices
(SLs).  Moreover, the semiconductor SLs in which it was first observed were
the MnTe/CdTe \cite{gieb1}, MnTe/ZnTe \cite{rhyne} and EuTe/PbTe
\cite{gieb2}, all with  antiferromagnetic layers. Recently, 
such coupling was also identified in semiconductor multilayer structures
with ferromagnetic EuS, \cite{kepa,ha} and Ga(Mn)As layers \cite{akiba}. While
the IEC in the trilayer Ga(Mn)As/Ga(Al)As/Ga(Mn)As with the high
concentration of free carriers, \cite{akiba}, can be, at least
qualitatively, explained in terms of the models tailored for metallic
systems, the other semiconductor structures exhibiting IEC call for a
different approach.

Several attempts to explain IEC in all-semiconductor structures have
already been
reported in the literature. Two models, in which the interlayer coupling is
mediated by carriers localized on shallow impurities in the spacer region,
were proposed for II-VI SLs,\cite{shevch, rusin}. 
These models do not apply to IV-VI structures, with the PbTe
and PbS spacers, since in lead chalcogenides localized shallow impurity
states were never detected \cite{heinrich}. For these SLs, mechanisms of
interlayer spin-spin interactions mediated by valence-band electrons were
suggested. The calculations of the difference between total electronic
energies obtained for two different (B1 and B2 in Fig.~\ref{a-fm}) 
spin configurations of the SL,
performed within a frame of a very simple 1D tight-binding model, put first
in evidence the significant role of the valence band electrons in IEC in
all-semiconductor magnetic/nonmagnetic layer structures \cite{blin}. This
role was further demonstrated for EuTe/PbTe/EuTe trilayers by Wilczy\'nski
and \'Swirkowicz in \cite{swir}, where a 3D tight binding model, still
oversimplifying the band structure, was used. A different approach to the
magnetic interlayer interactions mediated by valence electrons have been
chosen by Dugaev et al, \cite{dugaev}.  These authors studied the
Blombergen-Rowland mechanism within the effective mass approximation and
obtained a ferromagnetic coupling between two magnetic impurities situated
at the opposite interfaces of a narrow-gap IV-VI semiconductor spacer. As
the experimentally observed IEC in EuS/PbS SLs is antiferromagnetic,
\cite{kepa, ha}, this means that the Blombergen-Rowland interactions 
are 
not dominating IEC in these SLs. Still, because of the low concentration of
free carriers and the absence of shallow impurity states in PbS, the IEC in
these SLs is most likely mediated by valence electrons.  In this situation,
the total energy calculations, which do not focus on a particular
interaction mechanism but account globally for the spin-dependent structure
of valence bands, seem to constitute the most appropriate approach. The
calculations of this type, reported in \cite{blin} and \cite{swir}, were clearly
oversimplified and performed for a different spin structure than that of
EuS/PbS (001) SL.

In this paper we present the results of refined 3D  total energy 
calculations, which take
into account the crystal and the band structures of the SLs' component 
materials.  The tight-binding model with its  assumptions and the 
results for three different spin structures, corresponding
to all experimentally studied IV-VI semiconductor magnetic/nonmagnetic SLs,
are presented in Section III.  In Section II the magnetic and band
structures of the constituent materials are described and the SL geometry
is specified.  The comparison with experimental data and the conclusions
are presented in the last Section.
       
\section{Constituent materials and the superlattices geometry}

All the components of the EuX/PbX SLs, where X=Te or S, crystallize in the
rock-salt structure.  Bulk PbS and PbTe are narrow gap nonmagnetic
semiconductors with very similar band structures.  They both have a direct
energy gap between the p-anion valence band and the p-cation conduction
band at the point L of the BZ \cite{wei}. It is well known that in both
these lead chalcogenides the spin-orbit terms are important for the
detailed description of the energy bands in the vicinity of the L
point,\cite{lin}. One notices, however, that the spin-orbit corrections
affect predominantly the conduction bands originating from the states of
the heavy Pb atoms.
 
Bulk EuS is a classical Heisenberg ferromagnet with the Curie temperature
16.6 K \cite{wachter}.  Bulk EuTe exhibits type II AFM
structure with the N\'eel temperature 9.6 K
\cite{wachter}. In EuTe the spins of Eu ions are ferromagnetically ordered 
in (111)-type planes, 
which in turn are coupled antiferromagnetically to one another.  All the Eu
chalcogenides are semi-insulating, large gap semiconductors.  The results
of the nonrelativistic APW calculations of the EuS spin-polarized band
structure \cite{cho} show the narrow f($\uparrow$) bands situated in the
energy gap between the valence band, formed essentially of anion p states,
and the conduction band, built mostly of cation d states. The valence band
maximum is situated at the center of the Brillouin zone (BZ) and the
conduction band minimum at the point X. The spin splitting of the valence
band results predominantly from the spin dependent mixing of p anion and f
cation states, whereas that of the conduction band is mostly due to f-d and
s-d on-site direct exchange. Much less is known about the EuTe band
structure. The optical experiments performed at T=300~K indicate
\cite{guntherod}, that in the paramagnetic phase of EuTe
the f-d gap is somewhat larger than in EuS, in
agreement with the general trends visible in the experimental data \cite{wachter} and
the results of APW
calculations for other europium chalcogenides \cite{cho}.
These trends seem not to be followed in the recent \cite{jaya} calculations of EuTe band 
structure, focused predominantly on the conduction bands.
 
\begin{figure}
\includegraphics*[width=70mm]{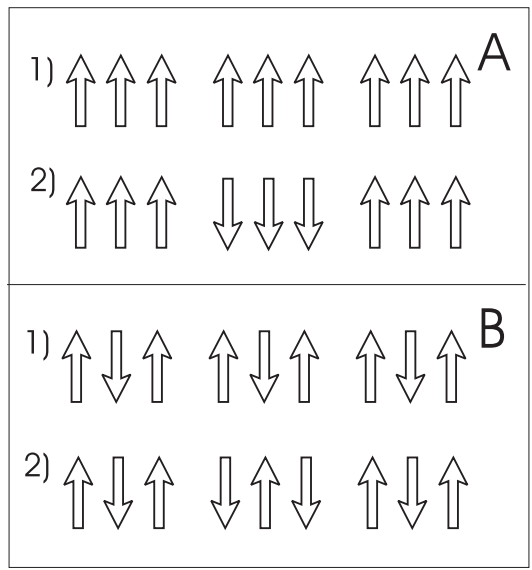}
\caption[]{The correlated co-linear spin structures for (A) ferromagnetic 
and (B) antiferromagnetic layers. For the "in-phase" spin structures A1 and
B1 the magnetic period is equal to the chemical one, for the "out-of-phase"
spin structures A2 and B2 the magnetic period is two times larger.}
\label{a-fm}
\end{figure}

\begin{figure}
\includegraphics*[width=85mm]{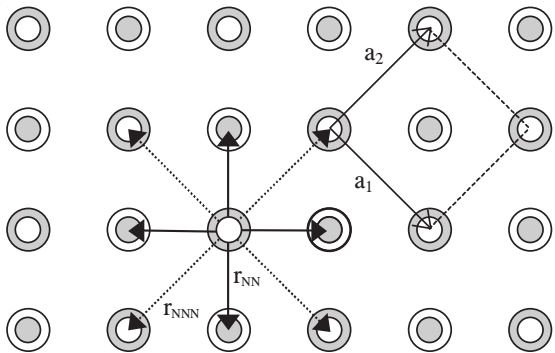}
\caption[]{Schematic view of two successive atomic layers of the EuS/PbS SL 
grown along [001] axis. The solid gray circles represent anions, the white
circles cations -- small circles are for ions in upper layer, the big ones
for ions in the other. The distances r$_{NN}$ from a cation to the four
in-plane NN anions are shown by solid lines, by dotted lines the distances
r$_{NNN}$ to the in-plane NNN cations}
\label{001}
\end{figure}

Two types of EuS/PbS SLs were experimentally studied, one grown on KCl
substrate along [001]
and the other on BaF$_2$ along the [111] crystallographic axis. 
The measurements show that in both cases 
the ferromagnetically ordered Eu spins within each magnetic layer
lie in the planes perpendicular to the growth axis \cite{stachow}.
In the (001) structures each atomic 
monolayer consists of both anions and cations, with the monolayers 
$a/2$ apart, where $a$ is the cubic lattice constant. The schematic 
view of two such monolayers is presented in  Fig.~\ref{001}.
The distances between the cation and its four in-plane nearest
neighbors (anions) and four in-plane next nearest neighbors (cations) are
also shown in the figure. The spin structure of the magnetic (001) SL,
which corresponds to  the observed antiferromagnetic interlayer
coupling is shown schematically in Fig.~\ref{a-fm}, A2.

\begin{figure}
\includegraphics*[width=85mm]{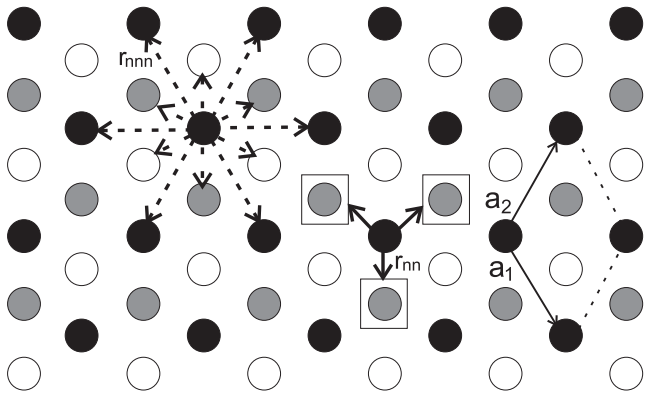}
\caption[]{Schematic view of the crystallographic structure of the EuX/PbX 
(111) SLs. The ions in three successive cation planes are represented by 
gray, black and white circles, respectively. Only three NN anions, lying 
in the layer $a\sqrt{3}/6$ above the black cation plane, are shown by 
open squares. The distances to all twelve NNN are marked by dotted lines}
\label{111}
\end{figure}

In the case of (111) EuS/PbS SLs IEC has
not been yet observed. These SLs have the same crystallographic structure
as the experimentally studied EuTe/PbTe SLs grown along the [111]
axis.  With this growth direction, the subsequent (111)
atomic monolayers are built either of anions or of cations, in
alternation. The distance between cation and anion monolayers is 
$a\sqrt{3}/6$. A schematic view of three successive cation layers is presented
in Fig.~\ref{111} - the analogous anion sublattice is shifted by
$a\sqrt{3}/2$ along the [111] direction and is 
not shown in the figure for the clarity reasons.

In EuTe/PbTe SLs the neutronographic measurements show that 
the AFM Type II structure is preserved in each EuTe
layer, but the FM spin sheets form exclusively on the (111) planes parallel
to the layers. This is in contrast to bulk EuTe, where there are four
symmetry-equivalent Type II AFM arrangements in which the ferromagnetic
spin sheets form on the \{111\}, \{$\overline{1}$11\},
\{$1\overline{1}$1\}, or \{$11\overline{1}$\} plane families. Moreover, for
the nonmagnetic spacers thin enough, the satellites observed in
neutronographic spectra clearly indicate the existence of some long range
order proving that the spins in consecutive magnetic layers are not
randomly oriented, but tend to align along the same direction in a
correlated way \cite{gieb2}. Although for the antiferromagnetic layers the
notions of AFM and FM IEC are not applicable, two types of co-linear
correlated spin orientations in successive layers are still possible:
identical (in-phase) and reversed (out-of-phase), as shown in
Fig.~\ref{a-fm}B. Both types of IEC were observed in the experiment.

All the (EuX)$_{m}$/(PbX)$_{n}$ SLs (where m and n denote the number of EuX
and PbX layers, respectively) have relatively small lattice mismatch.  This
mismatch, as well as the strains resulting from it, will be ignored in the
following, though the strains were shown to affect the magnetic properties
of the EuX layers, and their transition temperatures to the paramagnetic
state,\cite{stachow, goncha}.

To discuss the spin coupling between the magnetic layers one has to
consider a SL magnetic elementary cell containing at least two such layers.
In SLs grown along [001] crystallographic direction the situation is
simple, as the stacking sequence is ABAB-type.  Such stacking does not
enlarge the size of the magnetic elementary cell, whatever $m+n$ value.  In
contrast, the stacking sequence ABCABC-type (compare Fig.~\ref{111}) 
for both anions and cations,
in SLs grown along [111] axis does enlarge the elementary cell when $m+n$
is not a multiple of 3. Thus, to limit the size of elementary cells, we
consider only the (111) SLs with $(n+m)/3=integer$.  In the case of (001)
SLs the primitive lattice vectors, which define our elementary cells are:
${\bf{a_1}}=a/2[1,-1,0]$; ${\bf{a_2}}=a/2[1,1,0]$; ${\bf{a_3}}=a[0,0,m+n]$.
For
(111) SLs, when the z-axis is chosen along the growth direction, we have
${\bf{a_1}}=a/4[\sqrt{2},-\sqrt{6},0]$;${\bf{a_2}}=a/4[\sqrt{2},\sqrt{6},0]$;
${\bf{a_3}}=(2a/\sqrt{3})[0,0,m+n]$ (the plane views of these cells are
sketched in Fig.~\ref{001} and Fig.~\ref{111}).  For both types of SL, our 
magnetic
elementary cells contain, therefore, $2(m+n)$ anions, $2m$ magnetic cations
and $2n$ nonmagnetic cations.

In order to determine IEC in the above structures we compare the total
valence-electron energies in two magnetic SLs: with the in-phase and
out-of-phase spin ordering. For the n and m values typical for the
experimentally studied SLs the elementary cells contain several tens of
atoms. In view of this complexity, we decide to use the simplest
calculation scheme still leading to fairly realistic band structure,
namely, an empirical tight-binding method. Even though the one-electron
methods are not designed for the total-energy calculations, the small
spin-dependent changes in the total energy should be described adequately
within this approach.

\section{Tight-binding model}
To construct the empirical tight-binding Hamiltonian matrix one has to
select the set of atomic orbitals for every type of involved ions and 
to specify the
range of the ion-ion interactions.  This selection is always a compromise
between the best description of the band structure and the minimization of
the Hamiltonian matrix dimensions and of the number of parameters used. In
the following we assume that the proper description of SL band structure is
reached, when the Hamiltonian reproduces in the $n=0$ and $m=0$ limits the
known band structures of the bulk constituent magnetic and nonmagnetic
materials, respectively. This criterion determines in principle the
selection of the ionic orbitals and gives the values of the parameters, all
but those characterizing the interaction between magnetic and nonmagnetic
cations.

To calculate the band structure of the lead chalcogenides (PbS and PbTe) we
took into account s and p orbitals for both anions and cations, which lead
to 8x8 Hamiltonian matrix. We allowed for the s--s, s--p and p--p
anion-cation nearest-neighbor (NN) interactions and the anion--anion and
cation--cation p--p next-nearest neighbor (NNN) interactions.  It turned
out that the band structure can be reproduced much better when we include
also, by second--order perturbation, the interactions of p-orbitals with
the three NN d-orbitals belonging to the $F_2$ representation. The values
of the parameters describing all these interactions and the values of the
on-site orbital energies were determined by a $\chi^2$ minimization
procedure, in which the band structure was fitted to the energies in the
high symmetry points of the BZ, taken from \cite{lin} and \cite{wei}. The
obtained energy bands for PbS and PbTe along the symmetry axes of the BZ,
are presented in Fig.~\ref{pbtes}.
\begin{widetext}
\begin{center}
\begin{figure}[h]
\includegraphics*[width=0.8\textwidth]{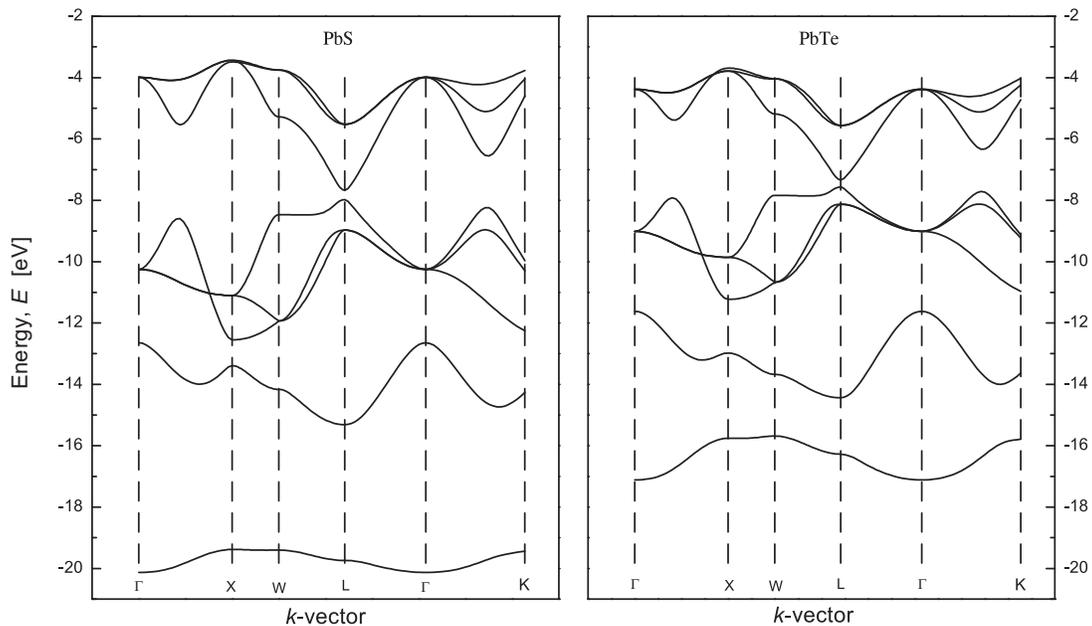}
\caption[]{Model band structures of PbS and PbTe}
\label{pbtes}
\end{figure}
\end{center}
\end{widetext}
 
In the other limit, for europium chalcogenides (EuS and EuTe), to describe
the cations we take explicitly the one s- and five d-orbitals,
whereas the anions are described as before by s- and p-orbitals. The NN
interaction involving the anion p-orbitals and cation s- and d-orbitals as
well as the NNN cation-cation d--d and anion-anion p--p interaction were
included in the 10x10 Hamiltonian. The s-anion--s-cation, s-anion--d-cation
interactions, turned out to be less important and were neglected. Instead,
we included, again by second order perturbation, the hybridization of anion
p($\uparrow$)-orbitals with the cation f$(\uparrow)$-orbitals - this was
necessary for reproducing the spin splittings of the valence bands in the
ferromagnetic EuS (we neglected the hybridization with the energetically
distant f$(\downarrow)$ band).  To reproduce the spin splittings in the EuS
conduction bands the on-site exchange constants $J_s$ and $J_d$ had to be
introduced.  The band structure of EuS, presented in Fig.~\ref{eutes} was 
obtained with the parameters fitted to the results of the APW spin--polarized
calculations reported in \cite{cho}. The presented in Fig.~\ref{eutes} 
band structure of EuTe was obtained 
with the parameter values extrapolated from the values for EuS and EuSe by
exploiting the chemical trends in europium chalcogenides. The elementary
cell of the antiferromagnetic EuTe has a twice larger volume and completely
different shape than the one of the ferromagnetic EuS -- to facilitate the 
comparison
with EuS, we present the band structure of EuTe in the paramagnetic 
phase.

In the above, the number of independent fitting parameters was partially
reduced according to the Harrison relations \cite{harrison}, e.g., instead
of two NNN interatomic matrix elements $pp\sigma$ and $pp\pi$, we used
$pp\sigma=-4 pp\pi$.  It has to be also noted, that in all calculations we
neglected the spin--orbit terms, known to be important in lead
chalcogenides. These terms would increase the number of model parameters
and double the matrix dimensions and would, therefore, pose a problem in
the case of the SLs. Fortunately, we are mainly interested in valence
bands, for which the spin--orbit is much less important than for the
conduction bands.
In the magnetic elementary cell of the 
(EuX)$_{m}$/(PbX)$_{n}$ SL there are
$4(m+n)$ nonequivalent ions. Seven orbitals (s,p,d) have to be taken into
account for each anion and each nonmagnetic cation and thirteen (s,d, and
f) for each magnetic cation. In principle, the SL tight-binding Hamiltonian
is therefore a $(40m+28n)$x$(40m+28n)$ matrix, which has to be completed by
the nonmagnetic cation--magnetic cation interactions. The constants
describing these interaction can not be inferred from the $m=0$ and $n=0$
limits, they were estimated from Harrison's formula for interatomic matrix
elements.  The matrix was then perturbationally reduced to the
$(20m+16n)$x$(20m+16n)$ matrix.  To determine the small difference between
the large total energies of the valence electrons in the two, in-phase and
out-of-phase, spin configurations, we did not calculate these energies
separately. Instead, the difference between the two energies of the valence
electrons was calculated at a given k-point, after the numerical
diagonalization of the two corresponding Hamiltonian matrices and summation
of the occupied states' energies. The results, obtained for a mesh of
k-points (16000 points in the case of the tetragonal BZ of the (100) SLs,
and 17280 in the case of the hexagonal BZ of the (111) SLs), were then used
in the triple Simpson procedure for integrating over the entire BZ.  
The calculations were
performed for (EuS)$_m$/(PbS)$_n$ SLs grown in both [001] and [111]
directions and for (EuTe)$_m$/(PbTe)$_n$ SLs grown along the [111] axis.

\begin{widetext}
\begin{center}
\begin{figure}[h]
\includegraphics[width=0.8\textwidth]{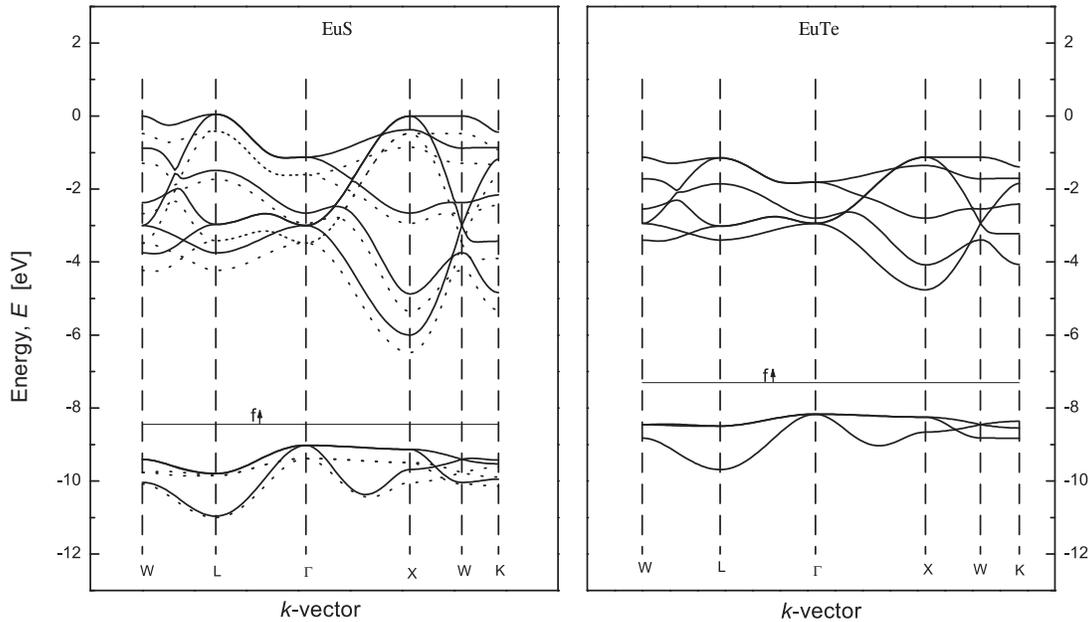}
\caption[textwidth=140mm]{
Model band structures of ferromagnetic EuS 
(solid lines represent
the spin-down bands; dotted lines -- spin-up bands) and paramagnetic EuTe
}
\label{eutes}
\end{figure}
\end{center}
\end{widetext}

We denote by ${\Delta}E$ the absolute value of the energy difference 
between the in-phase and out-of-phase spin configuration per
unit surface of the layer. ${\Delta}E$ can be regarded as a measure of the
strength of the interlayer spin coupling in the SLs -- for ferromagnetic
structures it can be expressed in terms of the constant $J_1$ \cite{parkin}, 
commonly used to characterize the IEC in metallic magnetic/nonmagnetic 
trilayers by the relation: ${\Delta}E=4 |J_1|$ (here
the factor 4, instead of 2, accounts for the fact, that in SL each magnetic
layer is coupled to two neighboring layers). 

The sign of the calculated energy difference determines the spin configuration
in consecutive magnetic layers. In the ferromagnetic, EuS-based, SLs the
out-of-phase spin configuration in consecutive magnetic layers is
energetically preferred, so that IEC in these structures has an
ANTIFERROMAGNETIC character, in agreement with the experiment. 
For the antiferromagnetic, EuTe-based SLs the
situation is more complicated: for odd number $m$ of spin planes in the
magnetic layer the out-of-phase configuration has the lower energy, whereas
for even $m$ it is the in-phase configuration, which is  energetically
favored. Thus, one can notice that in all studied SLs the
valence electron mediated IEC prefers the spin configuration with 
the opposite directions of spins at the two interfaces bordering the
spacer. 

Many various $m$ and $n$ values were selected to study the range of the
interlayer coupling and the IEC dependence on the thickness of the magnetic
layer. It turned out, that in all SLs for fixed spacer thickness $n$ the
strength of IEC is almost independent on the magnetic layer thickness $m$.
This seems to prove that in the considered here SLs, composed of 
two semiconductors with very different energy gaps, 
the valence electron mediated IEC is essentially a surface
effect.  

\begin{figure}
\includegraphics*[width=85mm]{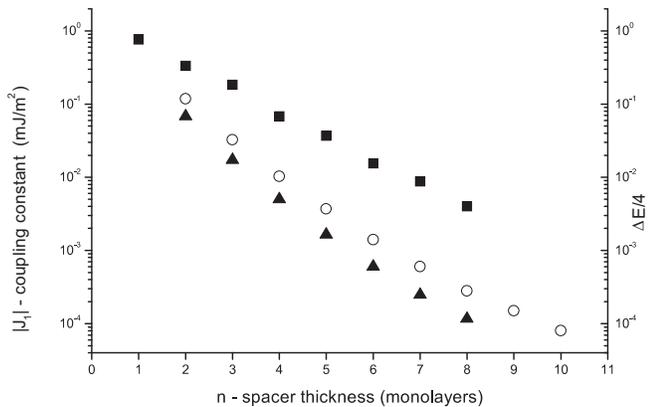}
\caption[]{The interlayer exchange constant $J_1$ (left Y-axis) as a function
of the spacer thickness for ferromagnetic EuS/PbS (001) (squares) and (111)
(triangles) SLs. For the antiferromagnetic EuTe/PbTe SLs (circles) the
absolute value of the energy difference $\Delta$$E$ was divided by 4 (at the
right Y-axis) for comparison with the FM case}
\label{J1}
\end{figure}

The dependence of the strength of the interlayer coupling on the
spacer thickness $n$ for all three studied types of SLs is
presented in Fig.~\ref{J1}. We recall that for the [111] growth direction, we
calculated IEC only for $(n+m)/3=integer$, so that for these SLs in the
figure the points for different $n$ values do not correspond to the same
$m$. As can be seen in Fig.~\ref{J1}, the strength of the coupling in all 
three cases decreases with the spacer thickness approximately
exponentially. 
The strongest and the least rapidly decreasing IEC was obtained in the case
of the FM EuS/PbS (001) SLs. The comparison of the results obtained with
the same set of model parameters for the two different types of EuS/PbS SLs,
(i.e., the grown along [001] and [111] axis) 
indicates that the valence electron mediated IEC depends strongly on the
lattice geometry. For example, for the PbS spacer $n=2$ the obtained coupling 
between EuS (111) magnetic layers is about five times weaker than that
between EuS (001) layers. Moreover, in the (111) SLs the strength of the
coupling decreases more quickly with $n$.
The small regular deviations from the smooth dependence of $J_1$ vs. $n$, 
which can be best seen for every second $n$ in the (001) case, but also
for every third $n$ in the (111) results, reflect the periodic effects
of stacking.

The IEC calculated for the AFM EuTe/PbTe (111) SL turned out to
be stronger than that for FM EuS/PbS (111) SL (see Fig.~\ref{J1}).
The difference in the band parameters  by itself does not explain this 
result -- the calculations performed for AFM and FM (111) SLs with
identical sets of band parameters have shown that the coupling between the
AFM layers is approximately two times stronger. This indicates that the
valence electron mediated IEC is not a purely surface effect. 
In view of the present results for IEC in antiferromagnetic EuTe/PbTe SLs,
we note that the simple 1D model in \cite{blin} overestimated  
the strength of IEC for large spacer thicknesses -- now the decrease of 
IEC with 
$n$ is much more rapid; the character of the interlayer coupling 
was however described properly. 

\section{Comparison with the experimental data}

\subsection{Ferromagnetic EuS/PbS superlattices}
The IEC was observed in ferromagnetic EuS/PbS (001) SLs by magnetic \cite{ha} 
and neutron
diffraction and reflectivity \cite{kepa} methods. For the samples with the
thin enough spacers, i.e., 4.5$\AA$ (1.5 monolayers, probably a mixture of
$n=2$ and $n=1$) and 10$\AA$ (ca $n=3$), the antiferromagnetic interlayer
coupling was observed in the magnetic moment measurements; the magnetic
methods did not reveal any IEC in the samples with larger spacer
thicknesses. In the neutronographic experiments the AFM IEC was confirmed
in the above two samples, but it was also observed in the sample with
23$\AA$ PbS layers. Further measurements in the external magnetic fields,
parallel to the layers, allowed to estimate the experimental strength of the
coupling constant $J_1$, from the magnetic field $B$ erasing the AFM
neutron reflectivity peak \cite{kepa2}. This was possible for the two
samples with the thinner spacers, but not for the sample with the thicker
spacer. For the latter, the field-induced changes in the neutronographic
spectra were irreversible, what suggests that in this case the IEC was
weaker than the magnetic anisotropies.  The estimated experimental values
of $J_1$ are: 0.063, 0.031 and 0.019 (in mJ/m$^2$), for n=1, 2 and 3,
respectively, \cite{kepa2}.  
The corresponding theoretical values obtained from our model
are: 0.77, 0.33, 0.18 mJ/m$^2$. Thus, one can conclude, that the model of
valence electron mediated IEC describes properly the sign and the rate of
the decrease of the coupling with the spacer thickness, but overestimates the strength of the coupling. The fact that the 
theoretical results obtained for crystallographicaly perfect SLs lead to 
exchange constants order of magnitude larger than those observed 
for the real multilayer structures, is probably due to the interface
diffusion, which in the case of metallic structures was shown to reduce 
significantly the strength of the IEC \cite{bruno}. 

\subsection {Antiferromagnetic EuTe/PbTe superlattices}
Unfortunately, for the AFM type of SLs there are no experiments, 
which provide direct information about the strength of the coupling. 
The evidences of the existence of the coupling between the
AFM EuTe layers come from the satellite structure of the
neutron diffraction spectra, seen in a variety of EuTe/PbTe SLs
consisting of several hundreds of periods,\cite{gieb2, icps}. 
The detailed 
analysis of the shapes of the satellite lines in the neutronographic spectra
 indicates that in these SLs the EuTe layers are not entirely coupled, but
only partially correlated, the less the thicker are the PbTe spacer layers
of the SL \cite{icps}. 
Under the strong assumption that the structures are morphologically perfect, 
with the same $m$ and $n$ values throughout the entire SL, this degree of 
correlation can be quantitatively determined. Under the same assumption,
the analysis of the satellites positions allows to distinguish which spin
configuration, the in-phase or out-of-phase, is dominating.
 
The observed spectra for the SLs with nominally even $m$ and even $n$
reveal the preference for the in-phase spin configurations, whereas those
for the SLs with odd $m$ and even $n$ exhibit the preference for the
out-of-phase configuration, both in agreement with the predictions of the
present model. For the case of even $m$ and odd $n$ there are no
available data.  Finally, for the samples with $m$ and $n$ both odd, 
the neutron diffraction
spectra seem to indicate that the in-phase configuration is preferred,
contrary to the theoretical predictions. However, the in-phase spin
configuration for SLs with odd number of spin planes in each
antiferromagnetic layer should exhibit ferrimagnetic properties, i.e., lead
to a significant net magnetic moment of ferrimagnetic domains.  No such
magnetic moments were detected in these samples \cite{holl}. These somewhat
confusing results seem to indicate that both chemical and magnetic
structures of the studied SLs are not perfect enough. New technological and
experimental efforts to observe the IEC in EuTe/PbTe SLs with smaller
number of SL periods, i.e., in SLs with better controlled $m$ and $n$
values, are undertaken.

In conclusion, we have shown, within a 3D tight binding model, that the
valence electron mediated interlayer exchange coupling explains the AFM
coupling between the FM layers observed in EuS/PbS (001) SLs with narrow
PbS spacers. The strength of the calculated coupling depends strongly on
the lattice geometry and decreases approximately exponentially with the
spacer thickness $n$. For a given type of SL it is almost independent on
the number $m$ of the spin planes within each magnetic layer. These
features distinguish the considered mechanism from another mechanism of AFM
coupling between the FM layers, namely the dipolar coupling possible in
multilayer structures with tiny magnetic domains \cite{borchers}.

The valence electron mediated interlayer coupling is, up to now, the only
effective mechanism capable to explain the origin of the interlayer
correlations observed in the antiferromagnetic EuTe/PbTe SLs with no
localized impurity states. The current, not complete understanding of the
experimental data for AFM SLs, does not allow, however, to draw definite
conclusions about the comparison between the details of the experimental
and theoretical results.

Acknowledgments:

This work was partially supported by the Polish Scientific Committee KBN,
grant No. 2 P03B 007 16.

\end{document}